\documentclass[a4paper, 11pt]{article}
\usepackage{graphicx}
\usepackage{authblk}
\begin{document}


\title{Carbon macromolecules in the cycle of interstellar matter : observations and
laboratory experiments}

\author[]{Christine Joblin}
\affil[]{Centre d'Etude Spatiale des Rayonnements, CNRS-UPS-OMP, BP 4346, 31028 Toulouse Cedex 04, France}
\date{}
%

\maketitle
\begin{abstract} Carbon macromolecules are intermediates between small gas-phase species
and larger dust structures.  I illustrate how observations and dedicated laboratory
experiments support this picture.
\end{abstract}
%
\section{Astrophysical context}
The Aromatic Infrared emission Bands (AIBs) between 3.3 and 12.7\,$\mu$m are well observed
in  UV-irradiated regions. The identity of their carriers is not precisely established but
polycylic aromatic hydrocarbons (PAHs) remain the best candidates.
Energetic budget calculations have led to estimates of 10 to 20\% of total carbon contained
in these species, making them an important component of interstellar matter.
Trying to match the AIBs with the spectra of laboratory samples does not appear
to be a successful method to identify interstellar PAHs. 
One has to think where these species come from and how they evolve
due to environmental conditions (UV irradiation, gas and dust interactions).
Several topics can be addressed:\\
{\bf -i-} PAHs contain hydrogen at their periphery. Photodissociation can affect
the hydrogenation coverage. Is this a mechanism to produce H$_2$? \\
{\bf -ii-} Is the fragmentation of the carbon skeleton of PAHs a reservoir to produce small gas-phase species? \\
{\bf -iii-} What is the connection between PAHs and larger structures (the VSGs of D\'esert et al. 1990)?\\
{\bf -iv-} PAHs can play an important role in chemistry. They offer a large surface and can contain 
some very active dangling bonds.\\
As the AIBs are the only evidence for the presence of PAHs, infrared spectroscopy is the obvious
diagnostic to answer these questions. Since dependence on physical conditions (UV field and
gas density) is searched for, extended photodissociation regions (PDRs) are good objects to be
observed providing additional imaging capabilities. Examples are given in the next section.
The astronomical media considered are characterized by low densities, typically less than
10$^{6}$\,cm$^{-3}$ and cold temperatures, a few tens of K for PAHs between the absorption of
two UV photons. PIRENEA is an experimental set-up that was developed according to these
specifications (high isolation and low temperatures) by combining the trapping capabilities of an ion
cyclotron resonance (ICR) cell with cryogenic shielding. Some recent results obtained with PIRENEA
are presented in section 3.
 
\section{Questions raised by observations}
Items (i) and (iv) invoke an evaluation of the role of PAHs in the formation of H$_2$.
The spatial distribution of the emission in the AIBs has been compared by several authors
with that in the H$_2$ ro-vibrational lines in the near-IR (Brooks et al. 2000,
Habart et al. 2003, Joblin et al. 2003a). 
Comparison with the predictions of PDR models underlies the difficulty to account for
the H$_2$ emission observed in regions where the AIBs are bright. It has therefore been proposed
that PAHs in these regions probably contribute to the formation of H$_2$
(Habart et al. 2003). The various mechanisms that can take place have been reviewed by
the authors: reaction of H atoms (free or physisorbed) with chemisorbed H (Cazaux \& Tielens
2002) or photodissociation by UV photons (Joblin et al. 2000).\\
The question of the coupling of PAHs with small hydrocarbons (item -ii-) has recently gained
some interest thanks to the detection  with radiotelescopes of species such as C$_2$H,
c-C$_3$H$_2$, C$_4$H and C$_6$H  in PDRs  (Foss\'e 2003). The observed abundances
cannot be accounted for by PDR
models (Gerin et al 2003). Could the photodestruction of PAHs replenish the gas-phase
in small hydrocarbons or carbon clusters?
Another major question is that of the formation of PAHs. Where do these molecules so abundant
in PDRs come from? The analysis of data on reflection nebulae provided by the camera CAM
on board the European Infrared Space Observatory brings us some new clues. Bright AIBs
are observed at the cloud interfaces but an additional continuum appears
to rise when going inside the cloud (Abergel et al. 2002). Using a singular value decomposition
analysis, a full spectrum was extracted for the additional emitting species (Boissel et al. 2001,
Rapacioli et al. these proceedings). There is a continuum rising towards longer wavelengths but
also associated bands which are generally broader than the AIBs but keep some similarities.
Furthermore, the decrease in the emission of this component at the border of the cloud is found
to correlate with the increase in the emission of the AIBs. This led the authors to propose that
the species carrying the continuum emission are VSGs formed of PAH clusters which evaporate
at the border of clouds and release free PAHs (cf. Rapacioli et al.). This sets item -iii-.

\begin{figure}[h]
   \centering
\includegraphics[width=12cm]{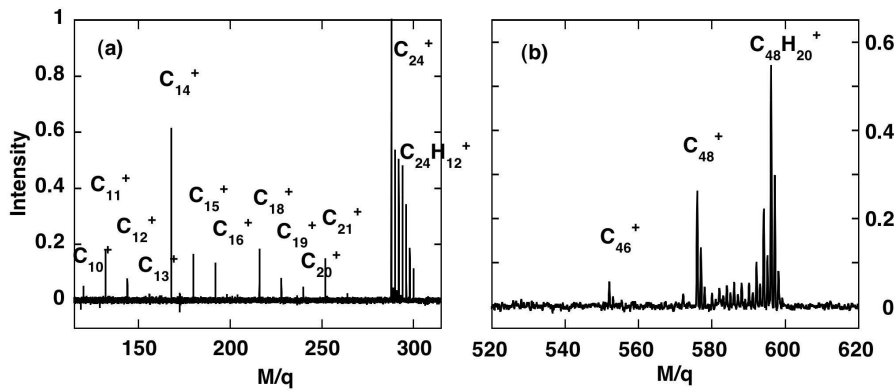}
      \caption{ Photodissociation of (a) $^{12}$C$_{24}$H$_{12}^+$  and (b) C$_{48}$H$_{20}^+$
    (includes $^{13}$C isotopomers) under continuous visible UV irradiation: (a) 120 s of Xe lamp
    + OPO laser tuned at 480 nm and (b) 30 s of Xe lamp + OPO laser tuned at 440 nm.}
       \label{figure_mafig}
   \end{figure}
 
\section{Laboratory approach}
Questions raised by observations can be addressed in the laboratory. The PIRENEA set-up was
specifically developed for studies related to large molecules such as PAHs or very small grains
of astrophysical interest. A description of the experiment can be found in
Joblin et al. (2001). At that time, the aim was to optimize the photophysical  interface. It was
shown that the photodissociation of the coronene PAH cation $^{12}$C$_{24}$H$_{12}^+$
under continuous irradiation of a Xe lamp leads to preferential formation of even-mass species
$^{12}$C$_{24}$H$_{2n}^+$ with n=[0,5]. It was not clear whether dehydrogenation was proceeding
through H$_2$ formation or sequential loss of H, the odd-mass species being fragile
intermediates easily destroyed under continuous irradiation. Using the capability of
selective ejection provided by the ICR cell, more recent experiments confirm the latter
explanation (Joblin et al. 2003b).\\
The C$_2$H$_2$ channel was found to be negligible in the photodissociation of the compact PAH
C$_{24}$H$_{12}^+$, whereas it is known to be a major channel in the fragmentation of small
species n$\leq$14 (Jochims et al. 1994). For C$_{24}$H$_{12}^+$, the destruction of the carbon skeleton occurs only
after complete dehydrogenation. The fragmentation of C$_{24}^+$  leads to a series of carbon
clusters (Fig.\,1). Since the experiment is performed under continuous radiation, some of them
are nor primary daughter species but produced by further fragmentation. Calculations by
Jones \& Seifert (1997) provide the various isomeric forms of C$_{24}$, the highest thermodynamic stability
being attained by the coronene skeleton, closed cage fullerene  and ring structures including
the bicyclic ring 14C-10C. The presence of the latter form in our experiment is strongly suggested by
the strong peak of C$_{14}^+$ (Fig.\,1). C$_{11}^+$ is then formed by loss of  C$_3$.
Looking at the spectrum in Fig.\,1, it also appears that there are other dissociation paths involving
different isomers of C$_{24}^+$ and a dissociation cascade producing C, C$_2$ and C$_3$. Although
a specific study would be necessary for each species, the global picture appears
to be consistent with previous studies  (cf. for instance Shelimov et al. 1994; Pozniak \& Dunbar 1997). 
First results were also obtained on the larger molecule dicoronene C$_{48}$H$_{20}^+$. In this case
again, the photodissociation proceeds until full dehydrogenation followed by loss of C$_2$ by the pure
carbon cluster C$_{48}^+$ (Fig.\,1b). This is the usual dissociation channel for fullerenes, the prototypical example
being C$_{60}$ (cf. review by Lifshitz 2000).\\
Results have just been obtained with PIRENEA on the reactivity of PAH cations
(including dehydrogenated species) with various gases, O$_{2}$ and H$_{2}$O in particular.
Studies dedicated to PAH clusters are planned in the near future.\\
\\
{\bf Acknowledgments}\\
M. Armengaud and P. Frabel are acknowledged for their technical involvment in the PIRENEA
set-up, the Programme National de Physique et Chimie du Milieu Interstellaire and the
R\'egion Midi-Pyr\'en\'ees for their financial support. 
Special thanks to M. Gerin and D. Teyssier for involving me in observations of small
hydrocarbons in PDRs and to J. Cernicharo for encouraging the studies on carbon clusters.


\end{document}